\documentclass[11pt]{article}
\usepackage{amsmath, amsthm, amssymb}

\input xy
\xyoption {all}

\newcommand{\ket}[1]{| #1 \rangle}
\newcommand{\bra}[1]{\langle #1|}
\newcommand{\ip}[2]{\langle #1|#2 \rangle}
\newcommand{\bracket}[3]{\langle #1|#2|#3 \rangle}

\newcommand{\be}{\begin{equation}}
\newcommand{\ee}{\end{equation}}
\newcommand{\bea}{\begin{eqnarray}}
\newcommand{\eea}{\end{eqnarray}}
\newcommand{\bes}{\begin{equation*}}
\newcommand{\ees}{\end{equation*}}
\newcommand{\beas}{\begin{eqnarray*}}
\newcommand{\eeas}{\end{eqnarray*}}

\newtheorem{thm}{Theorem}[section]
\newtheorem{cor}[thm]{Corollary}
\newtheorem{lem}[thm]{Lemma}

\newtheorem{defn}[thm]{Definition}

\begin{document}

\title{Quantum walks on directed graphs}

\author{Ashley Montanaro\footnote{
Department of Computer Science, University of Bristol, Woodland Road, Bristol, BS8 1UB, England. e-mail: montanar@cs.bris.ac.uk} }

\maketitle

\begin{abstract}
We consider the definition of quantum walks on directed graphs. Call a directed graph {\em reversible} if, for each pair of vertices $(v_i, v_j)$, if $v_i$ is connected to $v_j$ then there is a path from $v_j$ to $v_i$. We show that reversibility is a necessary and sufficient condition for a directed graph to allow the notion of a discrete-time quantum walk, and discuss some implications of this condition. We present a method for defining a ``partially quantum'' walk on directed graphs that are not reversible.
\end{abstract}


\section{Introduction}

Random walks play an important r\^ole in classical computer science, so it seems plausible that a quantum counterpart could be equally important in the study of quantum computation. Quantum walks on undirected graphs have been defined using two different formulations (discrete-time \cite{aharonov} and continuous-time \cite{farhi}), and are known to exhibit markedly different behaviour to classical random walks \cite{kempe, childs}. Quantum walks have been used to produce novel quantum algorithms \cite{shenvi, childs2, ambainis} displaying speed-ups over their classical equivalents. A natural question arises: can a quantum walk be defined on a {\em directed} graph? If so, which directed graphs allow a reasonable definition?

As motivation for this, there are many problems in graph theory that are known or suspected to be more difficult to solve for directed graphs than undirected graphs (an example being {\sc Reachability} \cite{papadimitriou}, c.f.\ section \ref{reachability} below). It is interesting to ask whether quantum walk algorithms can provide any improvement over classical algorithms for such problems.

The continuous-time formulation of quantum walks works by introducing a quantum system whose Hamiltonian is based on the adjacency matrix of the graph. This will not be suitable for walks on directed graphs, as this matrix will not in general be Hermitian, and hence the evolution of the system will not be unitary. Therefore, this note will only consider the discrete-time formulation, which consists of the iterated application of a unitary operator based on the structure of the graph.

We give a necessary and sufficient condition - which we term {\em reversibility} - on a graph for it to allow a meaningful definition of a discrete-time quantum walk. We then discuss the implications of this result. If a directed graph does not allow the definition of a ``fully quantum'' walk that preserves coherence throughout, we provide a method for defining a walk that alternates unitary evolution and measurement, and still allows for a level of coherence to be maintained.


\subsection{Quantum walks on graphs}

We begin with some standard graph-theoretic definitions that will be used throughout this paper.

\subsubsection{Graphs}

A {\em graph} (or {\em digraph}; we will use the terms interchangeably) $G$ is defined here as a set of vertices $V$ and arcs $A$, where $A$ is a set of ordered pairs of vertices. We assume that there may be at most one arc in each direction between each two vertices. An {\em undirected graph} has the further property $(v_i,v_j)\in A \Leftrightarrow (v_j,v_i)\in A$. When $(v_i,v_j)\in A$, we say that $v_i$ is {\em connected to} $v_j$, and use the notation $v_i \rightarrow v_j$. We say that $G$ is {\em connected} if for every pair of vertices $(v, w)$ there is a sequence of vertices $v_1,v_2,... v_k$ such that $v=v_1$, $w=v_k$, and each consecutive pair of vertices is connected by an arc (in either direction, which may vary along the sequence).

The {\em out-neighbours} of a vertex $v_i$ are the vertices to which $v_i$ is connected; similarly, the {\em in-neighbours} of $v_i$ are the vertices that are connected to $v_i$. The {\em in-degree} and {\em out-degree} of $v_i$ are the number of in-neighbours and out-neighbours it has, respectively. Every vertex in a {\em d-regular} graph has $d$ in-neighbours and $d$ out-neighbours. A {\em subgraph} $G'$ of a graph $G$ is a graph whose sets of vertices and arcs are subsets of those in $G$. A {\em connected component} of $G$ is a maximal connected subgraph of $G$. A {\em path} is an ordered list of vertices $\{v_1,v_2,...\}$ where $v_{i-1} \rightarrow v_i$, for all $i>1$. A {\em cycle} is a path whose final vertex is the same as its initial vertex.

The adjacency matrix of $G$ is the matrix also called $G$, where $G_{ij}=1 \Leftrightarrow j \rightarrow i$. The support of a matrix $U$ is the matrix $U'$, where $U'_{ij}=0$ if $U_{ij}=0$, and $U'_{ij}=1$ otherwise. The (di)graph of a unitary matrix $U$ is the graph whose adjacency matrix is the support of $U$.


\subsubsection{Quantum walks}
\label{walks}

A {\em coined quantum walk} on a $d$-regular undirected graph $G$, as defined in \cite{aharonov}, is produced by creating a Hilbert space $\mathcal{H}_v$ of dimension $|G|$ (where $|G|$ is the number of vertices in $G$), and identifying a basis state with each vertex. Each arc leaving a vertex is labelled by an integer from $0$ to $d-1$. This space is then augmented with a ``coin'' Hilbert space $\mathcal{H}_c$ of dimension $d$. A ``coin toss'' operator $C$ is defined, which operates only on $\mathcal{H}_c$. Also, a ``shift'' operator $S$ is defined such that $S\ket{c}\ket{v_i}=\ket{c}\ket{v_j}$, where $v_j$ is the vertex at the other end of the arc from $v_i$ labelled by $c$. One step of the walk then consists of applying the unitary $S(C\otimes I)$ - i.e.\ a coin toss followed by a shift. Multiple coins can be used to define a walk on an irregular graph.

We now define a more general notion of a discrete-time quantum walk, using a similar definition to \cite{aharonov}.

\begin{defn}
\label{walk}
A discrete-time quantum walk is the repeated application of a unitary operator $W$, where each application of $W$ is one step of the walk. To define a quantum walk on a graph $G$, we identify a finite set of one or more basis states $\{\ket{v_i^1}, \ket{v_i^2}, ...\}$ with each vertex $v_i$ of the graph. We say a quantum walk can be implemented on $G$ if there exists a $W$ such that, for all $i$, $j$, $v_i \rightarrow v_j$ if and only if there exist $k$, $l$ such that $\langle v_j^k|W|v_i^l \rangle \neq 0$. We assume that $G$ has self-loops at each vertex.
\end{defn}


\section{Reversible and irreversible graphs}

\begin{defn}
An arc $a \rightarrow b$ is called {\em reversible} if there is a path from $b$ to $a$. A graph whose arcs are all reversible is also called reversible; otherwise, it is called {\em irreversible}.
\end{defn}

Consider the following examples. A graph containing at least one source or sink is irreversible. All undirected graphs are reversible. An {\em Eulerian} graph is a graph whose every vertex has equal in-degree and out-degree. All Eulerian graphs are reversible, as they admit Eulerian tours (a cycle that visits every vertex, and traverses each arc once). Thus, all regular graphs are reversible. A {\em Cayley} graph is a graph associated with a group $X$, whose vertices are the elements of $X$, and which contains an arc $v_a\rightarrow v_b$ if and only if the associated element $b=ac$, for some $c\in X$. All Cayley graphs are regular, and hence are reversible.

\begin{figure}[htp] \[
  \begin{array}{c@{\hspace{1cm}}c@{\hspace{1cm}}c}
  \raisebox{-0.6cm}{\xymatrix{*{\bullet}\ar[r] & *{\bullet}}} &
  \raisebox{-0.6cm}{\xymatrix{*{\bullet}\ar[r] & *{\bullet} \ar@(r,u)[]}} &
  \xymatrix{
  *{\bullet}\ar[r]\ar@{-}[d] & *{\bullet}\ar@{-}[d]\\
  *{\bullet}\ar[r] & *{\bullet}}
  \end{array}
  \] \caption{Some irreversible graphs. An undirected edge denotes an arc in both directions.}
  \label{fig1}
\end{figure}
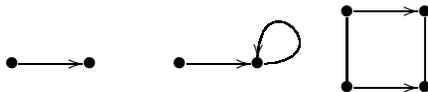

In the language of graph theory, every component of a reversible graph is {\em strongly connected} \cite{chen}. A reversible graph is almost the same as the transition graph of an irreducible Markov chain. However, there is a minor difference in that the definition here allows a graph to have multiple disconnected components, whereas irreducible Markov chains do not. So-called ``time reversible'' Markov chains are quite different, referring to a Markov chain which is symmetric in time \cite{kelly}.

The main result in this note is the following theorem. The proof will be given in section \ref{proof} below.

\begin{thm}
\label{reversibility}
A discrete-time quantum walk can be defined on a finite graph $G$ if and only if $G$ is reversible.
\end{thm}

\begin{cor}
The digraph of a unitary matrix is reversible.
\end{cor}

This corollary is simply the special case where each vertex of the graph is identified with one basis state.


\subsection{Determining reversibility}
\label{determine}

How easily can reversibility be determined for a given graph? On the one hand, it is clear that reversibility is a {\em global} attribute of a graph: it is not possible to determine whether a given arc is reversible without potentially considering all the other arcs in the graph. On the other hand, reversibility of a graph - or a given arc - can be determined in a time polynomial in the number of vertices in the graph. To see this, note that there is a simple algorithm (e.g.\ \cite{papadimitriou}) for determining whether there is a path between two vertices of a graph, which runs in time $O(n^2)$, where $n$ is the number of vertices. Using this algorithm to determine whether there is a path from each out-neighbour of each vertex $v_i$ back to $v_i$, reversibility of the entire graph can be calculated in time $O(n^4)$.


\section{Previous work}

A {\em directed bridge} is an arc in a graph $G$ whose removal would increase the number of connected components of $G$. Severini has proven \cite{severini2} that the digraph of a unitary matrix does not contain any directed bridges. The reversibility condition given here is stronger: a graph containing two connected components with multiple arcs between them, all going in the same direction, is irreversible. It has also been shown \cite{severini1} that the digraph of a unitary matrix is strongly quadrangular, and that it is without cut-vertices and bridges.

In terms of quantum walks, Severini has also shown that the underlying digraph of a coined walk is a line digraph \cite{severini3}. With the result given here, this implies that a line digraph is reversible. Acevedo and Gobron \cite{acevedo} have considered the classification of quantum walks on Cayley graphs. The study of discrete-time quantum walks was initiated by \cite{aharonov}.


\section{Proof of Theorem 2.2}
\label{proof}

\subsection{Necessity}

Our definition of a quantum walk consists of an identification of states with vertices of a graph. We will show that, if it is possible to ``walk'' from state $\ket{a}$ to state $\ket{b}$ by performing a unitary operation $W$, it is also possible to walk from $\ket{b}$ to $\ket{a}$ by performing $W$ a positive number of times. (Without this positivity condition, the problem is trivial, as multiplying by $W^{-1}$ will perform the reverse of $W$.)

\begin{lem}
\label{recurrence}
For any vector $\ket{a}$ in a finite-dimensional Hilbert space, any unitary operator $W$, and any $\epsilon > 0$, there exists $n\ge 1$ such that $|\bracket{a}{W^n}{a}| > 1-\epsilon$.
\end{lem}

\begin{proof}
This is simply a restatement in the terminology of unitary operators of the Quantum Recurrence Theorem proved by Bocchieri and Loinger in the language of wavefunctions \cite{bocchieri}, which in turn is a quantum equivalent of Poincar\'e's recurrence theorem for classical mechanics from 1890.
\end{proof}

The implication of this lemma is that repeating the same unitary operator enough times on $\ket{a}$ will produce a state arbitrarily close to $\ket{a}$.

\begin{lem}
\label{reverse}
For any vectors $\ket{a}$, $\ket{b}$ in a finite-dimensional Hilbert space, and for any unitary operator $W$, if $\bracket{b}{W}{a}\neq 0$, then there exists $m\ge 0$ such that $\bracket{a}{W^m}{b}\neq 0$.
\end{lem}

\begin{proof}
First, we have $\bracket{b}{W}{a}\neq 0 \Rightarrow \bracket{a}{W^{-1}}{b}\neq 0$. Consider a state close to $\ket{b}$, denoted by $\ket{b'}$. For sufficiently small $\epsilon$, $|\ip{b}{b'}|>1-\epsilon \Rightarrow \bracket{a}{W^{-1}}{b'}\neq 0$. By Lemma \ref{recurrence}, for arbitrarily small $\epsilon > 0$, there exists $p\ge 1$ such that $|\bracket{b}{W^p}{b}| > 1-\epsilon$. Set $\ket{b'}=W^p\ket{b}$ and we have $\bracket{a}{W^{-1}W^p}{b}\neq 0$, and hence $\bracket{a}{W^m}{b}\neq 0$ for $m=p-1\ge 0$.
\end{proof}

\begin{lem}
\label{path}
Let $W$ be a quantum walk defined on a graph $G$ with vertices $\{v_1, v_2, ...\}$ by associating basis states $\{\ket{v^1_i},\ket{v^2_i},...\}$ with each vertex $v_i$. Then, if for some $k$ and $l$, and $n\ge 0$, $\bracket{v^l_j}{W^n}{v^k_i}\neq 0$, there is a path from $v_i$ to $v_j$.
\end{lem}

\begin{proof}
$W^n\ket{v^k_i}$ describes $n$ steps of the walk starting from state $\ket{v^k_i}$, and hence produces a superposition over possible paths of length $n$ that the walk can take from vertex $v_i$. If $\bracket{v^l_j}{W^n}{v^k_i}\neq 0$ for some $k$, $l$, this implies that at least one of these paths must reach vertex $v_j$. For the case $n=0$, $\bracket{v^l_j}{W^0}{v^k_i}\neq 0$ only if $v_i=v_j$, as expected.
\end{proof}

Lemma \ref{reverse} shows that, if there is some amplitude to travel from some basis state $\ket{v^k_i}$ to some basis state $\ket{v^l_j}$ after 1 step of the walk, there must also be some amplitude to travel from $\ket{v^l_j}$ to $\ket{v^k_i}$ after $m\ge 0$ steps of the walk. With Lemma \ref{path}, this shows that, if there is an arc from the corresponding vertex $v_i$ to $v_j$, then there is a path from $v_j$ to $v_i$, and hence the necessity of Theorem \ref{reversibility} is proven.


\subsection{Sufficiency}
\label{sufficiency}

We will show that a coined quantum walk can be defined on any reversible graph. As defined in section \ref{walks}, we will use a Hilbert space $\mathcal{H}_v$ which associates one basis state with each vertex of the graph, augmented with a ``coin'' space $\mathcal{H}_c$. Our construction will be determined by the cycles in the graph.

\begin{lem}
\label{cycle}
Every arc in a reversible graph $G$ is included in at least one cycle.
\end{lem}

\begin{proof}
Let $v_i\rightarrow v_j$ be any arc in $G$. Since $G$ is reversible, there is a path from $v_j$ to $v_i$, and hence there exists a cycle that includes the given arc.
\end{proof}

This shows that it is possible to find a set $\{c_1,c_2,...\}$ of cycles in $G$ such that every arc in $G$ is included in at least one cycle. Each cycle $c_i$ gives rise to a permutation $P_i$ as follows. If a vertex $v$ is in the cycle with arc $v\rightarrow v'$, then $P_i(v) = v'$; otherwise, $P_i(v)=v$.

We then associate one coin basis state with each permutation, and select a coin operator $C$ as in the standard definition of a coined quantum walk. The Grover diffusion matrix is a popular choice of coin operator \cite{moore}.

The quantum walk operator $W$, which operates on $\mathcal{H}_c \otimes \mathcal{H}_v$, can then be expressed as

\be W = \left(\sum_i \ket{i}\bra{i}\otimes P_i \right) (C \otimes I) \ee

This proves that reversibility is a sufficient condition for the definition of a quantum walk. As an example of this construction, consider the following directed graph with labelled vertices. (Recall that self-loops are always included at every vertex.)

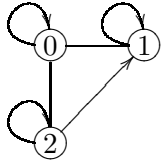
\bes
\xymatrix{*+[o][F-]{0} \ar@{-}[r] \ar@{-}[d] \ar@(l,u)[] & *+[o][F-]{1} \ar@(l,u)[] \\
*+[o][F-]{2} \ar[ur] \ar@(l,u)[]}
\ees

This graph admits the following four cycles, each augmented by self-loops at vertices not included in the cycle. Between them, these include every arc in the graph.

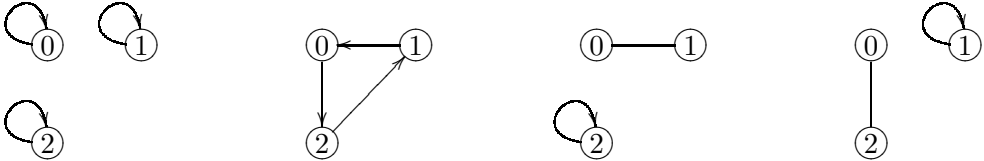
\bes
\xymatrix{*+[o][F-]{0} \ar@(l,u)[] & *+[o][F-]{1} \ar@(l,u)[] \\
*+[o][F-]{2} \ar@(l,u)[]}
\hspace{2cm}
\xymatrix{*+[o][F-]{0} \ar[d] & *+[o][F-]{1} \ar[l] \ar@{<-}[dl] \\
*+[o][F-]{2}}
\hspace{2cm}
\xymatrix{*+[o][F-]{0} \ar@{-}[r] & *+[o][F-]{1} \\
*+[o][F-]{2} \ar@(l,u)[]}
\hspace{2cm}
\xymatrix{*+[o][F-]{0} \ar@{-}[d] & *+[o][F-]{1} \ar@(l,u)[] \\
*+[o][F-]{2}}
\ees

We can now use a four-dimensional coin space to select between these four cycles. This example illustrates the fact that, depending on the structure of the graph in question, this algorithm may require a number of coin basis states exceeding the number of vertices in the graph. However, the number of coin states need never exceed the number of arcs. Also note that, for some graphs, the number of coin states used can be reduced by combining disjoint cycles into a single permutation.


\section{Simulating irreversible arcs with measurement}

There appears to be an intuitive correspondence between walking on a reversible graph and the reversibility of unitary evolution. Can we take this analogy a step further and define a quantum walk on a graph containing irreversible arcs by making use of the irreversible process of measurement? It turns out to be possible to define a ``partially quantum'' walk that maintains some quantum coherence in the reversible portions of the graph.

We will first define what is meant by ``reversible portions'' of a graph. Consider a subgraph $G'$ of a graph $G$. $G'$ is called a {\em reversible subgraph} of $G$ if, considered as a graph itself, $G'$ is reversible.

\begin{lem}
Let $G^{rev}$ be the subgraph of $G$ whose arcs consist of all the reversible arcs of $G$. Then $G^{rev}$ is reversible.
\end{lem}

\begin{proof}
For every arc $v_i\rightarrow v_j$ in $G^{rev}$, we require there to exist a path from $v_j$ to $v_i$. But this will be the case, because there is a path in $G$ from $v_j$ to $v_i$. Every arc in this path is reversible, and hence will be included in $G^{rev}$'s set of arcs.
\end{proof}

\begin{lem}
\label{partition}
It is possible to partition any graph $G$ into reversible subgraphs such that the arcs in $G$ that connect different subgraphs are all irreversible.
\end{lem}

\begin{proof}
Consider the connected components of $G^{rev}$, which are clearly reversible subgraphs of $G$. By definition, these do not contain any irreversible arcs. All the irreversible arcs in $G$ must therefore connect vertices in different reversible subgraphs of $G$.
\end{proof}

One possible way of defining a walk on an irreversible graph $G$ is the following approach. Informally, we consider $G$ as consisting of the connected components of $G^{rev}$ ``patched together'' with irreversible arcs. We produce a set of quantum walk operators, each corresponding to one component of $G^{rev}$. The irreversible arcs of $G$ are then simulated by replacing them with undirected edges. If such an edge is traversed by the ``walker'', we change to a different walk operator to ensure that it cannot be traversed in the opposite direction.

More specifically, consider vertices $v_1$ and $v_2$ that are in different reversible subgraphs of $G$ (called $C_1$ and $C_2$ respectively), and consider an irreversible arc $v_1 \rightarrow v_2$. This arc can be simulated by the following two-step process. First, we perform an incomplete measurement to determine whether the walker is in $C_1$ or $C_2$. Then, if it is in $C_1$, we perform one step of a quantum walk defined on the graph consisting of $C_1$ augmented with an undirected edge $v_1 \leftrightarrow v_2$. Alternatively, if the walker is in $C_2$, we perform one step of a walk only defined on the graph $C_2$. This ensures that the irreversible arc cannot be traversed in the wrong direction.

A more formal definition of this algorithm is given below.


\subsection{Algorithm to produce a partially quantum walk}
\label{algorithm2}

\begin{enumerate}
\item Determine which arcs in $G$ are irreversible (see section \ref{determine}), and thus produce a set of reversible subgraphs of $G$ (Lemma \ref{partition}).
\item Create a set of reversible graphs $\{G_1,G_2,...\}$ from the set of reversible subgraphs of $G$. These graphs partition all the vertices of $G$. Consider a Hilbert space $\mathcal{H}_v$ labelled by the vertices, and let $M$ be the incomplete measurement that projects onto this partition. Thus one measurement outcome corresponds to each reversible subgraph.
\item Consider each graph in turn, denoting the graph under consideration $G_i$. Some graphs $G_i$ will contain vertices that were the heads of irreversible arcs in $G$. Augment each graph $G_i$ with undirected links from these vertices to the corresponding targets of the arcs. Each of these links corresponds to moving to a new reversible subgraph. Call each augmented graph $G'_i$.
\item Define a coined quantum walk $W_i$ on each graph $G'_i$, using the approach of section \ref{sufficiency}.
\item We now have a set of quantum walks, each operating on a subgraph of the original graph. The overall walk consists of repeatedly alternating the measurement $M$ and one of the unitary walk operators. We perform measurement $M$, and if we see outcome $i$, we perform one step of the walk $W_i$.
\end{enumerate}

This approach has the advantage that it preserves coherence within each reversible subgraph; however, coherence across reversible subgraphs is not possible. That is, it is impossible to maintain a coherent superposition of states corresponding to vertices in two different subgraphs. An obvious implication of this is that a quantum walk on a graph whose arcs are all irreversible will be the same as the equivalent classical random walk.


\subsection{Example of the algorithm operating on an irreversible graph}

Consider the following labelled irreversible graph $G$ and its adjacency matrix. Self-loops are not shown here but should be considered to be present.

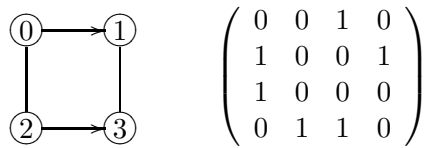
\bes
\raisebox{0.7cm}{\xymatrix{*+[o][F-]{0} \ar[r] \ar@{-}[d] & *+[o][F-]{1} \ar@{-}[d] \\
*+[o][F-]{2} \ar[r] & *+[o][F-]{3}}}
\hspace{1cm}
\left( \begin{array}{cccc}
0 & 0 & 1 & 0 \\
1 & 0 & 0 & 1 \\
1 & 0 & 0 & 0 \\
0 & 1 & 1 & 0
\end{array} \right)
\ees

We can split the graph into reversible subgraphs $R_1$ and $R_2$ consisting of the vertices $\{0, 2\}$ and $\{1, 3\}$, joined by irreversible arcs $0\rightarrow 1$ and $2\rightarrow 3$. These reversible subgraphs have adjacency matrices

\bes
R_1 = \left( \begin{array}{cccc}
0 & 0 & 1 & 0 \\
0 & 0 & 0 & 0 \\
1 & 0 & 0 & 0 \\
0 & 0 & 0 & 0
\end{array} \right) \mbox{~and~} R_2 =
\left( \begin{array}{cccc}
0 & 0 & 0 & 0 \\
0 & 0 & 0 & 1 \\
0 & 0 & 0 & 0 \\
0 & 1 & 0 & 0
\end{array} \right)
\ees

Define an incomplete projective measurement $M$ that distinguishes between $R_1$ and $R_2$. This measurement is made up of the operators
\bes
M_1 = \ket{0}\bra{0}+\ket{2}\bra{2} \mbox{~and~} M_2 = \ket{1}\bra{1}+\ket{3}\bra{3}
\ees

Then augment $R_1$ with undirected links corresponding to the irreversible arcs to $R_2$. This graph, denoted here by $R'_1$, is still reversible and allows the definition (omitted) of a coined quantum walk $W_1$. The subgraph $R_2$ does not need augmenting, as it does not contain the heads of any irreversible arcs, and a quantum walk $W_2$ can be defined on it directly.

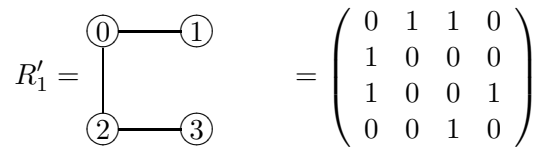
\bes
R'_1 =
\raisebox{0.7cm}{\xymatrix{*+[o][F-]{0} \ar@{-}[r] \ar@{-}[d] & *+[o][F-]{1} \\
*+[o][F-]{2} \ar@{-}[r] & *+[o][F-]{3}}}
\hspace{1cm}
= \left( \begin{array}{cccc}
0 & 1 & 1 & 0 \\
1 & 0 & 0 & 0 \\
1 & 0 & 0 & 1 \\
0 & 0 & 1 & 0
\end{array} \right)
\ees

A quantum walk on $G$ then consists of repeating the following steps. First, perform the measurement $M$ to determine whether the walker is in $R_1$ or $R_2$. If the measurement outcome is $M_1$, perform one step of the walk $W_1$ on the graph $R'_1$. Otherwise, perform one step of the walk $W_2$ on the graph $R_2$. Note that, if the walk is begun with a superposition corresponding to being at vertices 0 and 2, and outcome $M_2$ is measured after one step, this superposition is translated to a superposition of vertices 1 and 3 in $R_2$: quantum coherence is preserved.


\section{The Reachability problem for directed graphs}
\label{reachability}

{\sc Reachability} (also known as {\sc $st$-Connectivity} or {\sc Path}) is the problem of deciding whether, for two vertices $s$ and $t$ in a directed graph, there is a path from $s$ to $t$. In the context of classical algorithms, the problem is suspected to be more difficult than its undirected variant; in fact, it is NL-complete \cite{papadimitriou}, whereas undirected {\sc Reachability} is in L \cite{reingold}. On reversible directed graphs, the problem reduces to undirected connectivity. This is clear from the following lemma:

\begin{lem}
\label{connectivity}
In any connected reversible graph $G$, there is a path from every vertex $a$ to every other vertex $b$.
\end{lem}

\begin{proof}
Immediate from the definition of strong connectivity in \cite{chen}. To see this explicitly, note that since $G$ is connected, any vertex $a$ may be linked to any other vertex $b$ by a sequence of arcs $v_i \rightarrow v_j$ whose directions may vary along the sequence. Since $G$ is reversible, $v_i$ is also reachable from $v_j$ for each arc, so $a$ and $b$ are reachable from each other in either direction.
\end{proof}

Thus, every vertex within each connected component of a reversible graph is reachable from every other vertex in that component, exactly as in undirected graphs. Theorem \ref{reversibility} therefore implies that quantum walk algorithms may not be much help in solving {\sc Reachability}.

There are many other classical random walk algorithms which perform a search on directed graphs (an example being Sch\"oning's random walk algorithm for SAT \cite{schoning}). These often work by traversing a directed graph randomly until they reach a sink, which represents a previously unknown solution. Since such graphs are not reversible, the main result of this paper shows that quantum walk algorithms for such problems cannot be merely straightforward generalisations of their classical counterparts.

\subsection*{Acknowledgements}

I would like to thank Richard Jozsa for careful comments on this paper, and Sean Clark, Tobias Osborne and Tony Short for helpful discussions. This work was supported in part by the UK Engineering and Physical Sciences Research Council QIP-IRC grant.


\bibliographystyle{amsplain}

\end{document}